\documentclass[journal]{IEEEtran}
\usepackage{mathrsfs}
\usepackage{bbm}
\usepackage{amsmath,amscd,amssymb,latexsym}
\usepackage[all]{xy}
\usepackage{tabu}

\usepackage{graphicx}
\usepackage{xcolor}

\newtheorem{theorem}{Theorem}[section]
\newtheorem{lemma}[theorem]{Lemma}

\newtheorem{definition}[theorem]{Definition}
\newtheorem{example}[theorem]{Example}

\newtheorem{remark}[theorem]{Remark}

\ifCLASSINFOpdf
\else
\fi
\begin{document}
%
\title{New  LCD MDS codes of non-Reed-Solomon type}
%
%
%

\author{Yansheng~Wu,~
        Jong Yoon~Hyun,~
        and~Yoonjin~Lee~

\thanks{Manuscript received June 24, 2020; revised January 14, 2021; accepted May 08, 2021. Y. Wu was sponsored by NUPTSF (Grant No. NY220137),
J.Y. Hyun was supported by the National Research Foundation of Korea(NRF) grant
funded by the Korea government(MEST)(NRF-2017R1D1A1B05030707),
and Y. Lee by Basic Science Research Program
through the National Research Foundation of Korea(NRF) funded by the Ministry of Education (Grant No. 2019R1A6A1A11051177) and also
by the National Research Foundation of Korea(NRF) grant funded by the Korea government (MEST)(Grant No. NRF-2017R1A2B2004574).  (Corresponding author: Jong Yoon Hyun.) }

\thanks{Y. Wu is with the School of Computer Science, Nanjing University of Posts and Telecommunications, Nanjing 210023, China, and also with the  Shanghai Key Laboratory of Trustworthy Computing, East China Normal University, Shanghai, 200062,  China (e-mail: yanshengwu@njupt.edu.cn).}

\thanks{J.Y. Hyun is with the Konkuk University, Glocal Campus, 268 Chungwon-daero Chungju-si Chungcheongbuk-do 27478, South Korea (e-mail: hyun33@kku.ac.kr).}

\thanks{Y. Lee is with the Department of Mathematics, Ewha Womans
University, Seoul 120-750, South Korea (e-mail: yoonjinl@ewha.ac.kr).} 


}

%
%

\markboth{IEEE Transactions on information theory,~Vol.~, No.~, ~2021}{Wu \MakeLowercase{\textit{et al.}}:
New  LCD MDS codes of non-Reed-Solomon type}
%



\maketitle

\begin{abstract}
Both {\em linear complementary dual} (LCD) codes and {\em maximum distance separable} (MDS)
codes have good algebraic structures, and they have interesting practical applications
such as communication systems, data storage, quantum codes, and so on.
So far, most of LCD MDS codes have been constructed by employing {\it generalized Reed-Solomon codes}.
In this paper we construct some classes of {\it new} Euclidean LCD MDS codes and Hermitian LCD MDS codes
which are not monomially equivalent to Reed-Solomon codes, called LCD MDS codes {\it of non-Reed-Solomon type}.
Our method is based on the constructions of  Beelen et al. (2017) and Roth and Lempel (1989).
To the best of our knowledge, this is the first paper on the construction of LCD MDS codes of non-Reed-Solomon type; any LCD MDS code of non-Reed-Solomon type constructed by our method is not monomially equivalent to any LCD code constructed by the method of Carlet et al. (2018).
\end{abstract}


%
\IEEEpeerreviewmaketitle

\section{Introduction}
%
%
%
%
\IEEEPARstart{T}{he} concept of {\em linear complementary dual} (LCD) codes was introduced by
Massey in 1992 \cite{M}, and they have interesting applications in communication systems, cryptography,
and data storage. In particular, Carlet et al. \cite{BCCGM, CG2,CG1} found a new application
of binary LCD codes in implementations against side-channel and fault injection attacks.
Since then, LCD codes have attracted great attention from researchers in coding community.
Yang and Massey \cite{YM94} showed  a necessary and sufficient condition for a cyclic code
to guarantee LCD property as well. The authors of \cite{LDL2017,LLD2017} constructed
two families of both LCD and BCH codes.

A {\em maximum distance separable} (MDS) code has the greatest error correcting capability
when its length and dimension is fixed. MDS codes are extensively used in communications (for example,
 Reed-Solomon codes are all MDS codes), and they have good applications in minimum storage codes and quantum codes.  There are many known constructions for MDS codes; for instance, {\em Generalized Reed-Solomon} (GRS) codes \cite{RS}, based on the equivalent problem of finding $n$-arcs in projective geometry \cite{MS}, circulant matrices \cite{RL}, Hankel matrices \cite{RS2}, or extending GRS codes.



Both LCD codes and MDS codes have good algebraic structures, and they have interesting practical applications as mentioned above. Until now, constructions of most of LCD MDS codes have been achieved by using generalized Reed-Solomon codes (GRS codes) since GRS codes are all MDS. We discuss recent research progresses on LCD MDS codes as follows.

(1) Jin \cite{J} and Shi et al. \cite{SYY} constructed several LCD MDS codes using generalized Reed-Solomon codes with some additional conditions on special polynomials. 

(2) Chen and Liu \cite{CL} made a different approach to obtain some LCD MDS codes from generalized Reed-Solomon codes, and he extended some results by Jin \cite{J}. Afterwards, Luo et al.  \cite{LCC} and Fang et al. \cite{FFLZ} further extended the results of Chen and Liu and investigated Euclidean and Hermitian hulls of MDS codes and their applications in quantum codes.

(3) Beelen and Jin \cite{BJ} found an explicit construction of several LCD MDS codes in the odd characteristic case using the theory of algebraic function fields.

(4) Sari and  Koroglu \cite{SK} constructed MDS negacyclic LCD codes of some special lengths, and they provided lower bounds on the minimum distance of these codes.

(5) Carlet et al. \cite{CMTQ} obtained many parameters of Euclidean and Hermitian LCD MDS codes by using some linear codes with small dimension or codimension, self-orthogonal codes and generalized Reed-Solomon codes. 

(6) Carlet et al. \cite{CMTQP} introduced a general construction of LCD codes from any linear codes. 
More exactly speaking, if there is an $[n, k, d]$ linear code  over $\mathbb{F}_q$ $(q >3)$ (respectively, over $\mathbb{F}_{q^2}$ $(q>2)$), then there exists an $[n, k, d]$ Euclidean (respectively, Hermitian) LCD code over $\mathbb{F}_q$  (respectively, over $\mathbb{F}_{q^2}$).

\vskip 0.3cm

In this paper, we construct some classes of {\it new} Euclidean LCD MDS codes and Hermitian LCD MDS codes
which are not monomially equivalent to Reed-Solomon codes, called LCD MDS codes {\it of non-Reed-Solomon type}. To the best of our knowledge, this is the first paper on the construction of LCD MDS codes of non-Reed-Solomon type.
In the coding theory, it is an important issue to find all inequivalent codes of the same parameters. We point out that any LCD MDS code of non-Reed-Solomon type constructed by our method is not monomially equivalent to any LCD code constructed by the method of Carlet et al.~\cite{CMTQP}.
In particular, we construct some {\it twisted} Reed-Solomon codes or {\it Roth-Lempel} codes which are also LCD MDS codes of non-Reed-Solomon type; these codes cannot be constructed by the method in \cite{CMTQP}.
We also present some examples of non-Reed-Solomon LCD MDS codes, which are obtained by using our results and Magma implementation.

Our method is based on the constructions of  Beelen et al. \cite{BPR} and Roth and Lempel \cite{RL}.
For construction of the LCD MDS codes of non-Reed-Solomon type, first we use some special matrices to form a generator matrix such that the product of the generator matrix and its (conjugate) transpose is as simple as possible. Secondly, we need to add some conditions to make sure that the product is nonsingular. Finally, we use the lifting of the finite field so that these codes also have  the MDS property.

\vskip 0.3cm

This paper is organized as follows. In Section 2, we recall some basic concepts
on Euclidean and Hermitian LCD MDS codes and two constructions of these codes.
In Section 3, we find some new Euclidean and Hermitian LCD MDS codes, which are not monomially
equivalent to generalized Reed-Solomon codes. We finish this paper with a conclusion in Section 4.

\section{Preliminaries}

Let $\Bbb F_q$ be the finite field of order $q$, where $q$ is a power of an odd prime.
An ${[n, k]_q}$ linear code $\mathcal{C}$ over $\Bbb F_q$  is a $k$-dimensional
subspace of $\Bbb F_q^n$. The minimum distance $d$ of
a linear code $\mathcal{C}$ is bounded above by the so-called {\em Singleton bound}, that is, $d\le n-k+1$.
If $d= n-k+1$, then the code $\mathcal{C}$ is called a {\em maximum distance separable} (MDS) code.

For $x \in \Bbb  F_{q^2}$, the conjugate of $x$ is denoted by $\overline x = x^q$.
For a matrix $A$, we denote by $A^T$ the transpose of $A$, and $\overline A$ 
the matrix of conjugates of $A$. For a set $B=\{x_1,x_2,\ldots, x_l\}\subseteq \Bbb  F_{q^2}$, we define $\overline{B}=\{x_1^q, x_2^q, \ldots, x_l^q\}$.

 \subsection{Equivalence of codes}

We recall some equivalence notions of codes over the finite field $\Bbb F_q$ (see \cite[Sections 1.6 and 1.7]{HP}).

 \begin{definition}{\rm Let $\mathcal C_1$ and $\mathcal C_2$ be two linear codes of the same length over $\Bbb F_q$. Two linear codes $\mathcal C_1$ and $\mathcal C_2$ are {\it permutation equivalent} if there is a permutation matrix $P$ such that $G_1$ is a generator matrix of $\mathcal C_1$ if and only if $G_1P$ is a generator matrix of $\mathcal C_2$.
}
 \end{definition}

Recall that a \emph{monomial matrix} is a square matrix which has exactly one nonzero entry in each row and each column. A monomial matrix $M$ can be written either in the form of $DP$ or the form of $PD'$, where $D$ and $D'$ are diagonal matrices and $P$ is a permutation matrix.

\begin{definition}{\rm Let $\mathcal C_1$ and $\mathcal C_2$ be two linear codes  of the same length over $\Bbb F_q$,  and let $G_1$ be a generator matrix of $\mathcal C_1$. Then
 $\mathcal C_1$ and $\mathcal C_2$ are {\em monomially equivalent} if
 there is a monomial matrix $M$ such that $G_1M$ is a generator matrix of $\mathcal C_2$.
}
 \end{definition}

 \subsection{Euclidean and Hermitian LCD  codes}

Given a linear code $\mathcal C$ of length $n$ over $\Bbb F_q$ (resp. $\Bbb F_{q^2}$ ),
the Euclidean dual code  and the Hermitian dual code of $\mathcal C$
are defined by
$$\resizebox{9cm}{!}{ $\mathcal C^{\bot_ E}=\bigg\{(x_0, \ldots, x_{n-1})={\bf x}\in \Bbb F_q^{n} : \begin{array}{l}
\langle {\bf x},{\bf y}\rangle_{E}=\sum_{i=0}^{n-1}x_iy_i=0
 \\ ~  \forall~  {\bf y}= (y_0, \ldots,y_{n-1})\in \mathcal C\end{array}\bigg\} $} $$
 and $$\resizebox{9cm}{!}{ $\mathcal C^{\bot_ H}=\bigg\{(x_0, \ldots, x_{n-1})={\bf x}\in \Bbb F_{q^2}^{n} : \begin{array}{l}
\langle {\bf x},{\bf y}\rangle_{H}=\sum_{i=0}^{n-1}x_iy_i^q=0
 \\ ~  \forall~  {\bf y}= (y_0, \ldots,y_{n-1})\in \mathcal C\end{array}\bigg\},$} $$
respectively.

A linear code $\mathcal C$ over $\Bbb F_q$ is called a {\em Euclidean Linear Complementary Dual (Euclidean LCD)
code } if $\mathcal C \cap  \mathcal C^{\bot_ E} = \{0\}$, and it is called a {\em Hermitian Linear Complementary Dual (Hermitian LCD)
code} if $\mathcal C \cap  \mathcal C^{\bot_ H} = \{0\}$.

\begin{lemma}{\rm \label{lem1} \cite[Proposition 2]{CMTQ}
 If $G$ is a generator matrix of an $[n, k]_q$ linear code $\mathcal C$, then $\mathcal C$ is a Euclidean
(resp. Hermitian) LCD code if and only if the $k\times k$ matrix $GG^T$ (resp. $G\overline{G}^T$) is nonsingular. }

\end{lemma}

\subsection{Constructions of MDS codes}

In this subsection, we recall some developments on constructions of MDS codes, which include generalized Reed-Solomon codes, twisted  Reed-Solomon codes and Roth-Lempel codes as follows. Hereafter, we denote respectively by $G_1$ and $G_2$ 
the generator matrix of a twisted  Reed-Solomon code and a Roth-Lempel code.\\

We begin with the well-known generalized Reed-Solomon codes.

 \begin{definition}{\rm \label{def1} Let $\alpha_{1},\ldots,\alpha_{n}$ be distinct elements in $\Bbb F_{{q}} \cup \{\infty\}$  and $v_{1},\ldots,v_{n}$ be nonzero elements in $\Bbb F_{{q}}$. For $1\leq k \leq n$, the corresponding {\em generalized Reed-Solomon $(GRS)$ code} over $\Bbb F_{{q}}$ is defined by
 $$\resizebox{9cm}{!}{ $GR{S_k}({\boldsymbol{ \alpha}},{\bf v}): = \left\{({v_1}f({\alpha_1}), \ldots ,{v_{n }}f({\alpha_{n }})) \mid f(x) \in {\Bbb F_{q}}[x],\phantom{.} \deg(f(x)) < k \right\},$}$$
  where $\boldsymbol{ \alpha}=(\alpha_{1},\alpha_{2},\ldots,\alpha_{n})\in (\Bbb F_{{q}}\cup \{\infty\})^{n}$ and ${\bf v}=(v_{1},v_{2},\ldots,v_{n})\in {(\Bbb F_{{q}}^*)}^{n}$, and the quantity $f(\infty)$ is defined as the coefficient of $x^{k-1}$ in the polynomial $f$. 
}
 \end{definition}

If $v_i=1$ for every $i=1, \ldots n$, then $GR{S_k}({\boldsymbol{ \alpha}},{\bf v})$ is called a {\it Reed-Solomon $(RS)$ code.}
 It is well-known that a generalized Reed-Solomon code $GRS_k(\boldsymbol{\alpha},{\bf v})$ is an  $ {\left[ {n,k,n-k+1} \right] }$ MDS code. In fact,   $GRS_k(\boldsymbol{\alpha},{\bf v})$ has a generator matrix  as follows:  $$\resizebox{9cm}{!}{ $\left(\begin{array}{cclc} v_1 &v_2 &\ldots  &v_n\\ v_1\alpha_1 &v_2\alpha_2 &\ldots & v_n\alpha_{n}\\ \vdots &\vdots &\ddots&\vdots \\ v_1\alpha_1^{k-1} &v_2\alpha_2^{k-1} &\ldots & v_n\alpha^{k-1}_{n}\end{array}\right)=\left(\begin{array}{cclc} 1 &1 &\ldots  &1\\ \alpha_1 &\alpha_2 &\ldots &\alpha_{n}\\ \vdots &\vdots &\ddots&\vdots \\ \alpha_1^{k-1} &\alpha_2^{k-1} &\ldots & \alpha^{k-1}_{n}\end{array}\right) \left(\begin{array}{cclc} v_1 &0 &\ldots  &0\\ 0 &v_2 &\ldots &0\\ \vdots &\vdots &\ddots&\vdots \\ 0 &0 &\ldots & v_{n}\end{array}\right).$}$$



 In 2017, Beelen {\em et al.} \cite{BPR} presented a generalization of  Reed-Solomon codes, so-called {\it twisted Reed-Solomon codes}.

 \begin{definition} \label{def2} Let $\eta$ be a nonzero element in the finite field $\Bbb F_q$. Let $k,t$ and $h$ be nonnegative integers such that $0\le h<k\le q$, $k<n$, and $0<t\le n-k$.  Let $\alpha_{1},\ldots,\alpha_{n}$  be distinct elements in $\Bbb F_{{q}} \cup \{\infty\}$, and we write $\boldsymbol{\alpha}=(\alpha_{1},\alpha_{2},\ldots,\alpha_{n})$. Then the corresponding {\em twisted Reed-Solomon code} over $\Bbb F_q$ of length $n$ and dimension $k$ is given by \begin{eqnarray*}\resizebox{9cm}{!}{ $\mathcal C_k(\boldsymbol{\alpha}, t,h,\eta)=
\{(f(\alpha_1),  \cdots, f(\alpha_{n})): f(x)=\sum_{i=0}^{k-1}a_ix^i+\eta a_hx^{k-1+t}\in \Bbb F_q[x]\}$}.\end{eqnarray*}

 \end{definition}

 In fact,  \begin{equation}\label{eq1}\resizebox{8cm}{!}{ $G_1=\left(\begin{array}{cclcc}
1 &1 &\ldots  &1\\
 \alpha_1 &\alpha_2 &\ldots & \alpha_{n}\\
   \vdots &\vdots &\ddots&\vdots \\
   \alpha_1^{h-1} &\alpha_2^{h-1} &\ldots & \alpha^{h-1}_{n}\\
   \alpha_1^h+\eta \alpha_1^{k-1+t} &\alpha_2^h+\eta \alpha_2^{k-1+t} &\ldots & \alpha_{n}^h+\eta \alpha_n^{k-1+t}\\
  \alpha_1^{h+1} &\alpha_2^{h+1} &\ldots & \alpha^{h+1}_{n}\\
  \vdots &\vdots &\ddots&\vdots \\
  \alpha_1^{k-1} &\alpha_2^{k-1} &\ldots & \alpha^{k-1}_{n}\end{array}\right)$  } \end{equation}
 is the generator matrix of the twisted Reed-Solomon code $\mathcal C_k(\boldsymbol{\alpha}, t,h,\eta)$.

 

Note that in general, the twisted Reed-Solomon codes are not MDS. Beelen {\em et al.} \cite{BPR} obtained some results on the twisted Reed-Solomon codes as follows:

 \begin{lemma}\label{lem2} \cite[Theorem 17]{BPR} {\rm  Let $\Bbb F_s \subset \Bbb F_q$ be a proper subfield and $\alpha_{1},\ldots,\alpha_{n}\in \Bbb F_s$. If $\eta \in\Bbb F_q \backslash \Bbb F_s $, then the twisted Reed-Solomon code $\mathcal C_k(\boldsymbol{\alpha}, t,h,\eta)$ is MDS.}
 \end{lemma}

 As indicated  in \cite[Remark 8]{BPR2}, there is a mistake in the proof  of \cite[Theorem 18]{BPR}. Here we present an exact statement of \cite[Theorem 18]{BPR}. 
To do so, recall from \cite[Theorem 1]{BPR} that any MDS code having a generator matrix of  the form $[I_k \mid \mathbf{ A}]$,  is a GRS if and only if all  $3\times 3$ minor of $\widetilde{\mathbf{ A}}$ are zero, where $\mathbf{ A}=(A_{ij})$ and $\widetilde{A}_{ij}=A_{ij}^{-1}$. 

Let $\eta$ be in $\Bbb F_q^*$ such that the twisted Reed-Solomon code $\mathcal C_k(\boldsymbol{\alpha}, t,h,\eta)$ is MDS.
Let $[I_k \mid \mathbf{ A}]$ be a generator matrix of $\mathcal C_k(\boldsymbol{\alpha}, t,h,\eta)$.  Let $M_i=\frac{p_i(\eta)}{q_i(\eta)}$ for $i=1, 2, \ldots, l$ be all $3\times 3$ minors of $\widetilde{\mathbf{ A}}$, where $\mathbf{ A}=(A_{ij})$ and $\widetilde{A}_{ij}=A_{ij}^{-1}$. Here $p_i, q_i$ are polynomials over $\Bbb F_q$, and $q_i(\eta)\neq 0$. 

\begin{lemma}\label{lem3} \cite[Theorem 18]{BPR}    Let  $\alpha_{1},\ldots,\alpha_{n}\in \Bbb F_{{q}} $ and $2<k<n-2$.  Let $H=\{\eta\in\Bbb F_q^*: \mbox{the twisted Reed-Solomon code } \mathcal C_k(\boldsymbol{\alpha}, t,h,\eta) \mbox{ is MDS} \}.$ Assume that a certain $3\times3$  minor $M_i=\frac{p_i(\eta)}{q_i(\eta)}$ of $\widetilde{\mathbf{ A}}$ defined in right above is nonzero for some $\eta \in H$. 
Then $\mathcal C_k(\boldsymbol{\alpha}, t,h,\eta)$ for such  $\eta \in H$ is a non-Reed-Solomon code.
\end{lemma}  
   


 


 \begin{remark} (1) In Lemma \ref{lem3}, if $p_i(\eta)=0$ for any $3\times 3$  minor of $\widetilde{\mathbf{ A}}$, then  $\mathcal C_k(\boldsymbol{\alpha}, t,h,\eta)$ for  $\eta\in H$ is monomially equivalent to an RS code. By \cite[Corollary 2]{BPR}, an $[n,k,n-k+1]$ MDS code with $k<3$ or $n-k<3$, is monomially equivalent to an RS code. On the other hand, it was proved  in \cite[Corollary 9.2]{B} that  for a linear MDS code over $\Bbb F_{p^n}$ with parameters $[p^n+1, k(\le p), p^n-k+2]$ is an RS code. 
 
 (2)  To find a twisted MDS code of non-Reed-Solomon type, we need to check the minor assumption of Lemma \ref{lem3}.    $\blacksquare$

 \end{remark}
 

Combining Lemma \ref{lem2} and Lemma \ref{lem3}, we have 

  \begin{lemma} \cite[Corollary 20]{BPR}  Let $\Bbb F_s \subset \Bbb F_q$ and $\alpha_{1},\ldots,\alpha_{n}\in \Bbb F_s$.  Let $2<k<n-2$ and $n\le s$. Assume that the minor condition for $\eta\in\Bbb F_q \backslash \Bbb F_s$ of Lemma \ref{lem3} holds. 
Then $\mathcal C_k(\boldsymbol{\alpha}, t,h,\eta)$ is MDS but not monomially equivalent to an RS code.

 \end{lemma}


 \begin{remark}  In the next section, to obtain LCD MDS codes, we first study the LCD  property of  twisted  Reed-Solomon codes, and then we use a suitable vector $\boldsymbol{\alpha}$. Note that Lemma 2.6 shows the existence of MDS twisted  Reed-Solomon code, and in general, it is also hard to find an element $\eta\in \Bbb F_q$ such that $\mathcal C_k(\boldsymbol{\alpha}, t,h,\eta)$ is a non-Reed-Solomon LCD MDS code even for small lengths.


 \end{remark}

Roth and Lempel \cite{RL} found a new construction of MDS codes of non-Reed-Solomon type. A set $S\subseteq \Bbb F_q$ of size $m$ is called an $(m,t, \delta)$-set in $\Bbb F_q$ if there exists an element $\delta\in \Bbb F_q$ such that no $t$ elements of $S$ sum to $\delta$. We note that if $S$ belongs to some subfield $\Bbb F_s$ of $\Bbb F_q$, then the set $S$ is an $(m,t, \delta)$-set in $\Bbb F_q$ for each $\delta\in \Bbb F_q \backslash \Bbb F_s$.


\begin{definition} {\rm \cite{RL} Let $n$ and $k$ be two integers such that $k\ge 3$ and $k+1\le n\le q$.  Let $\alpha_1, \ldots, \alpha_{n}$ be distinct elements of $\Bbb F_q$, $\delta\in \Bbb F_q$, and $\boldsymbol{\alpha}=(\alpha_{1},\alpha_{2},\ldots,\alpha_{n})$. Then an $[n+2,k]$ {\em Roth-Lempel  code}
$RL({\boldsymbol{ \alpha}},k,n+2)$
over $\Bbb F_q$ is generated by the matrix \begin{equation}\label{eq2} G_2=
\left( \begin{array}{cccccc}
1 & 1& \ldots &  1 & 0 &0\\
\alpha_1& \alpha_2& \ldots &  \alpha_{n} & 0 &0\\
\vdots& \vdots& \ldots & \vdots & \vdots &\vdots\\
\alpha_1^{k-2}& \alpha_2^{k-2}& \ldots &  \alpha_{n}^{k-2} & 0 &1\\
\alpha_1^{k-1}& \alpha_2^{k-1}& \ldots &  \alpha_{n}^{k-1} & 1 &\delta\\
\end{array} \right).
\end{equation}
}
\end{definition}

\begin{lemma}{\rm \cite{RL} An $[n+2,k]$  Roth-Lempel  code with $k\ge 3$ and $k+1\le n\le q$
 over $\Bbb F_q$ is a non-Reed-Solomon code. Moreover, the Roth-Lempel  code is  MDS  if and only if the set $\{\alpha_1, \ldots, \alpha_{n}\}$ is an $(n, k-1, \delta)$-set in $\Bbb F_q$.

}

\end{lemma}


 \begin{remark} \label{lem6} {\rm  In the next section, in order to find LCD MDS codes,
 we first study  the LCD  property of   Roth-Lempel codes. Then we use some set $\{\alpha_1, \ldots, \alpha_{n}\}$ which is contained in some subfield of $\Bbb F_q$.
 }
 \end{remark}


Roth and Lempel \cite{RL} proved that the generator matrix $G_2$ in (2.2) cannot generate a GRS code by considering a certain form of codewords. That is, $G_2P$ can not generate a GRS code for any permutation matrix $P$.
By Definition 2.2, GRS codes are naturally monomially equivalent to RS codes. 
We claim that Roth-Lempel codes are not monomially equivalent to RS codes. 
Suppose that a  $RL({\boldsymbol{ \alpha}},k,n)$ code is monomially equivalent {\bf to}  a RS code. This means that there exists a monomial matrix $M$ such that $G_2M=G'$, where $G'$ is a generator matrix of a RS code. Recall that a monomial matrix $M$ can be written either in the form $DP$ or the
form $PD'$, where $D$ and $D'$ are diagonal matrices and $P$ is a permutation matrix.

 (1) If $M=DP$, then $G_2M=G_2DP=G'$. Hence, $G_2=G'P^{-1}D^{-1}$. Note that  $G'P^{-1}$  generates  some RS codes: that is, $G_2$ generates some GRS codes, which is a contradiction.

 (2) If $M=PD'$, then $G_2M=G_2PD'=G'$. Hence, $G_2=G'D'^{-1}P^{-1}$. We also see that $G_2$ generates some GRS codes; this leads to a contradiction.

In conclusion, Roth-Lempel codes are not monomially equivalent to RS codes. Moreover, in the whole paper, if a code is not monomially equivalent to an RS code, then we call it a code of {\em non-Reed-Solomon type} or a {\em  non-Reed-Solomon code}.

\section{New LCD MDS codes}

In this section, we use constructions in Section 2 to obtain new Euclidean and Hermitian LCD MDS codes. Subsections 3.1 and 3.2 deal with the Euclidean and Hermitian LCD MDS codes, respectively.

\subsection{Euclidean LCD MDS codes } 


Let $\gamma$ be a primitive element of $\Bbb F_q$ and $k\mid (q-1)$. Then $\gamma^{\frac{q-1}{k}}$ generates a subgroup of $\Bbb F_q^*$ of order $k$. Let $\alpha_i=\gamma^{\frac{q-1}{k}i}$ for $1\le i\le k$. One can easily check that  \begin{equation}\label{eq3}
\theta_f=\alpha_1^f+\cdots+\alpha_{k}^f=\left\{
\begin{array}{ll}
  k   &      \mbox{if}\ f\equiv0\pmod {k},\\
0 & \mbox{otherwise}.
\end{array} \right.
\end{equation}
Let

\begin{eqnarray}\label{eq4} \resizebox{8.1cm}{!}{ $ A_{\beta}=\left( \begin{array}{cccccc}
1 & 1& \ldots  & 1 &1\\
\beta\alpha_1& \beta\alpha_2& \ldots   &  \beta\alpha_{k-1}  & \beta\alpha_k \\
\vdots& \vdots& \ldots  & \vdots &\vdots\\
(\beta\alpha_1)^{k-1}& (\beta\alpha_2)^{k-1} &\cdots  & (\beta\alpha_{k-1})^{k-1} &(\beta\alpha_k)^{k-1} \\
\end{array} \right).$} 
\end{eqnarray}
 By Equation (\ref{eq3}), we have
\begin{eqnarray} \label{eq5}A_{\beta}A_{\beta}^T=\left( \begin{array}{ccccccc}
k &0&0& \ldots  & 0 &0\\
0& 0&0& \ldots   & 0  & \beta^k k \\
\vdots&\vdots& \vdots& \ldots  & \vdots &\vdots\\
0&0&\beta^k k&\cdots  & 0&0 \\
0&\beta^k k&0&\cdots  & 0&0 \\
\end{array} \right).
\end{eqnarray}
Let $C_{\beta}=A_{\beta}+B_{\beta}$, where $B_{\beta}$ is given by 
\begin{eqnarray}\label{eq6}\resizebox{8.5cm}{!}{$
\left( \begin{array}{cccccc}
0 & 0& \ldots  & 0 &0\\
\vdots& \vdots& \ldots  & \vdots &\vdots\\
\eta(\beta\alpha_1)^{k-1+t}& \eta(\beta\alpha_2)^{k-1+t}& \ldots   &  \eta(\beta\alpha_{k-1})^{k-1+t}  &\eta(\beta\alpha_k)^{k-1+t} \\
\vdots& \vdots& \ldots  & \vdots &\vdots\\
0 & 0& \ldots  & 0 &0\\
\end{array} \right)\begin{array}{l} \\   \\\leftarrow(h+1)th\\  \\ \\  \end{array}$}.\end{eqnarray}


The following lemma plays an important role in proving our main results of this Subsection 3.1.
First, we should find some conditions under which a twisted Reed-Solomon code $\mathcal C_k(\boldsymbol{\alpha}, t,h,\eta)$ over $\Bbb F_q$ becomes a Euclidean LCD code. 

\begin{lemma}\label{lem7}{\rm  Let $q$ be a power of an odd prime. If $k$ is a positive integer with $k\mid (q-1)$, $k<\frac{q-1}{2}$ and $h>0$, then there exists a $[2k,k]_q$ Euclidean  LCD twisted Reed-Solomon code $\mathcal C_k(\boldsymbol{\alpha}, t,h,\eta)$ over $\Bbb F_q$ for
$\boldsymbol{\alpha}=(\alpha_1,\ldots,\alpha_k,\gamma\alpha_1,\ldots,\gamma\alpha_k)$,
where $\gamma$ is a primitive element of $\Bbb F_q$ and $\alpha_i=\gamma^{\frac{q-1}{k}i}$ for $1\le i\le k$.

}

\end{lemma}

{\em Proof:}
By Definition \ref{def2}, to make sure that $\mathcal C_k(\boldsymbol{\alpha}, t,h,\eta)$ is a twisted Reed-Solomon code, we need $k\neq q-1$, and the entries of $\boldsymbol{\alpha}$ are all distinct, which are obvious. 
From Equation (\ref{eq1}), we recall that $G_1$ is a generator matrix of  the twisted Reed-Solomon code $\mathcal C_k(\boldsymbol{\alpha}, t,h,\eta)$ over $\Bbb F_q$.
It follows from Lemma \ref{lem1} that $\mathcal C_k(\boldsymbol{\alpha}, t,h,\eta)$ over $\Bbb F_q$ in Definition \ref{def2} is Euclidean LCD if and only if $G_1G_1^T$ is nonsingular.  Let $\theta_j=\sum_{i=1}^{n}\alpha_i^j$ and $l=k-1+t$. Then we compute $G_1G_1^T$ in  Equation (\ref{eq7}) in the top of next page.
\newcounter{mytempeqncnt}
\begin{figure*}[!h]
\normalsize
\setcounter{mytempeqncnt}{\value{equation}}
\setcounter{equation}{6} 
\begin{equation}
\label{eq7}
\resizebox{16cm}{!}{$G_1G_1^T = \left( \begin{array}{cccccccccccccccccccccccccccc}
n & \theta_{1}& \ldots  &\theta_{h-1}&\theta_{h}+\eta\theta_{l} & \theta_{h+1}  &\ldots& \theta_{k-2} &\theta_{k-1}\\
\theta_{1} & \theta_{2}& \ldots  &\theta_{h}&\theta_{h+1}+\eta\theta_{l+1} & \theta_{h+2}  &\ldots& \theta_{k-1} &\theta_{k}\\
\vdots & \vdots& \ldots  &\vdots&\vdots &  \vdots& \ldots  &\vdots&\vdots\\
\theta_{h-1} & \theta_h& \ldots  &\theta_{2h-2}&\theta_{2h-1}+\eta\theta_{l+h-1}& \theta_{2h}  &\ldots& \theta_{k+h-3} &\theta_{k+h-2}\\
\theta_{h} +\eta\theta_{l}& \theta_{h+1} +\eta\theta_{l+1}& \ldots  &\theta_{2h-1} +\eta\theta_{l+h-1}&\theta_{2h}+2\eta\theta_{l+h}+\eta^2\theta_{2l}& \theta_{2h+1}+\eta\theta_{l+h+1}& \ldots& \theta_{h+k-2}+\eta\theta_{l+k-2}&\theta_{k+h-1}+\eta\theta_{l+k-1}\\
\theta_{h+1} & \theta_{h+2}& \ldots  &\theta_{2h}&\theta_{2h+1}+\eta\theta_{l+h+1} & \theta_{2h+2}  &\ldots& \theta_{k+h-1} &\theta_{k+h}\\
\vdots & \vdots& \ldots  &\vdots&\vdots&   \vdots& \ldots  &\vdots&\vdots\\
\theta_{k-1} & \theta_{k}& \ldots  &\theta_{k+h-2}&\theta_{k+h-1}+\eta\theta_{l+k-1}&  \theta_{k+h}  &\ldots&\theta_{2k-3} &\theta_{2k-2}\\
\end{array}\right)$}\\
\end{equation}
\setcounter{equation}{\value{mytempeqncnt}}
\hrulefill 
\vspace*{4pt} 
\end{figure*}
\setcounter{equation}{2}


 Let $C_{\beta}=A_{\beta}+B_{\beta}$, where $A_{\beta}, B_{\beta}$ are given in Equation (\ref{eq4}) and Equation (\ref{eq7}), respectively. By Equation (\ref{eq5}), we have $C_{\beta}C_{\beta}^T$  in Equation  (\ref{eq8}) in the top of next page.
\begin{figure*}[!h]
\normalsize
\setcounter{mytempeqncnt}{\value{equation}}
\setcounter{equation}{7} 
\begin{equation}
\label{eq8}
\resizebox{16cm}{!}{$
C_{\beta}C_{\beta}^T
= \begin{pmatrix}
k &0& \ldots  & 0 &0\\
0& 0& \ldots   & 0  & \beta^k k \\
\vdots& \vdots& \ldots  & \vdots &\vdots\\
0&\beta^k k&\cdots  & 0&0 \\
\end{pmatrix} +
\begin{pmatrix}
0 & 0& \ldots  &0& \eta\beta^l\theta_{l}&0  &\ldots&0\\
0 & 0& \ldots  &0& \eta\beta^{l+1}\theta_{l+1}&0 &\ldots&0\\
\vdots & \vdots& \ldots  &\vdots& \vdots & \vdots& \ldots  &\vdots\\
0 & 0& \ldots  &0& \eta\beta^{l+h-1}\theta_{l+h-1}&0  &\ldots& 0\\
\eta\beta^l\theta_{l}& \eta\beta^{l+1}\theta_{l+1}& \ldots  &\eta\beta^{l+h-1}\theta_{l+h-1}&
2\eta\beta^{l+h}\theta_{l+h}+\eta^2\beta^{2l}\theta_{2l}&\eta\beta^{l+h+1}\theta_{l+h+1}&\ldots&\eta\beta^{l+k-1}\theta_{l+k-1}\\
0 &0& \ldots  &0& \eta\beta^{l+h+1}\theta_{l+h+1}&0 &\ldots&0\\
\vdots & \vdots& \ldots  &\vdots & \vdots & \vdots& \ldots  &\vdots\\
0 & 0& \ldots  &0 & \eta\beta^{l+k-1}\theta_{l+k-1}&0 &\ldots&0
\end{pmatrix}$}
\end{equation}
\setcounter{equation}{\value{mytempeqncnt}}
\hrulefill 
\vspace*{4pt} 
\end{figure*}
\setcounter{equation}{3}



Since every $\theta_t$ for $l\leq t\leq l+k-1$ is zero except exactly one $\theta_{t'}$, we
can rewrite (8) as
\begin{eqnarray*}&&C_{\beta}C_{\beta}^T=\left( \begin{array}{cccccc}
k &0& \ldots  & 0 &0\\
0& 0& \ldots   & 0  & \beta^k k \\
\vdots& \vdots& \ldots  & \vdots &\vdots\\
0&\beta^k k&\cdots  & 0&0 \\
\end{array} \right)\\
&+&\left( \begin{array}{cccccccc}
0 &\ldots&0& \ldots  & 0 &\ldots&0\\
\vdots&& \vdots& \ldots  & \vdots &&\vdots\\
0&\ldots &0& \ldots   & *_{\beta}  & \ldots&0 \\
\vdots&& \vdots& \ldots  & \vdots &&\vdots\\
0&\ldots &*_{\beta}& \ldots   & \Delta_{\beta}  &\ldots& 0 \\
\vdots&& \vdots& \ldots  & \vdots &&\vdots\\
0&\ldots&0&\cdots  & 0&\ldots&0 \\
\end{array} \right),\end{eqnarray*}
where $\ast_{\beta}$ and $\Delta_{\beta}$ are all elements in $\Bbb F_q$, the  $\ast_{\beta}$ and $\Delta_{\beta}$ are respectively entries located in the $(i+1,h+1)th$, $(h+1,i+1)th$ and $(h+1,h+1)th$ positions, and the other elements are all zero.





  Let $G_1=[C_1: C_{\gamma}]$. Then \begin{eqnarray*}&&G_1G_1^T=C_1C_1^T+C_{\gamma}C_{\gamma}^T\\
&=&\left( \begin{array}{cccccc}
2k &0& \ldots  & 0 &0\\
0& 0& \ldots   & 0  & (1+\gamma^k) k \\
\vdots& \vdots& \ldots  & \vdots &\vdots\\
0&(1+\gamma^k) k&\cdots  & 0&0 \\
\end{array} \right)\\
&+&\left( \begin{array}{cccccccc}
0 &\ldots&0& \ldots  & 0 &\ldots&0\\
\vdots&& \vdots& \ldots  & \vdots &&\vdots\\
0&\ldots &0& \ldots   & *_{1} + *_{\gamma}& \ldots&0 \\
\vdots&& \vdots& \ldots  & \vdots &&\vdots\\
0&\ldots &*_{1} + *_{\gamma}& \ldots   &  \Delta_1+\Delta_{\gamma}  &\ldots& 0 \\
\vdots&& \vdots& \ldots  & \vdots &&\vdots\\
0&\ldots&0&\cdots  & 0&\ldots&0 \\
\end{array} \right).\end{eqnarray*}

By the assumption on $k$ and $h>0$, we have $(\gamma ^k+1)k\neq0$. Hence, we can delete the element $*_{1} + *_{\gamma}$ by some elementary row and column operations of matrices  at the same time. Namely, we can  find an elementary matrix $P$ such that \\
\begin{eqnarray*}&&PG_1G_1^TP^T =
\left( \begin{array}{cccccc}
2k &0& \ldots  & 0 &0\\
0& 0& \ldots   & 0  & (1+\gamma^k) k \\
\vdots& \vdots& \ldots  & \vdots &\vdots\\
0&(1+\gamma^k) k&\cdots  & 0&0 \\
\end{array} \right)\\
&+&\left( \begin{array}{cccccccc}
0 &\ldots&0& \ldots  & 0 &\ldots&0\\
\vdots&& \vdots& \ldots  & \vdots &&\vdots\\
0&\ldots &0& \ldots   & 0& \ldots&0 \\
\vdots&& \vdots& \ldots  & \vdots &&\vdots\\
0&\ldots &0& \ldots   &  \Delta_1+\Delta_{\gamma}  &\ldots& 0 \\
\vdots&& \vdots& \ldots  & \vdots &&\vdots\\
0&\ldots&0&\cdots  & 0&\ldots&0 \\
\end{array} \right).\end{eqnarray*}
Since det$(P)=1$ and the element $ \Delta_1+\Delta_{\gamma}$ is the entry in the $(h+1)$th row and $(h+1)$th column, we have  det$(G_1G_1^T)=2(1+\gamma^k)^{k-1}k^k\neq 0$ by the definition of determinant.
Since the matrix $G_1G_1^T$ is nonsingular, the code $\mathcal C_k(\boldsymbol{\alpha}, t,h,\eta)$ is a Euclidean LCD code. This completes the proof. 
 $\blacksquare$

\begin{remark}
{\rm For $G_1=[C_1: C_{\gamma}]$, it is not guaranteed that $G_1G_1^T$ is nonsingular when $k=\frac{q-1}{2}$; if $\boldsymbol{\alpha}$ takes all non-zero elements of $\Bbb F_q$ and $1+\gamma^k=0$, then $G_1G_1^T$ is singular. However, if
 $q=5$ then we have $k=\frac{q-1}{2}=2$ and $G_1=\left( \begin{array}{cccccc}
1&1 & 1&1\\
2& 1&  2  & 0\\
\end{array} \right)$.
Then it is easy to verify that $\mathcal C_2(\boldsymbol{\alpha}, 1,1,1)$ is a Euclidean LCD code for $\boldsymbol{\alpha}=(1,2,-2,-1)$. 
}
\end{remark}

\begin{example}
{\rm
Let $q=3^4=81$, $k=4$, and $\gamma$ be a primitive element of $\Bbb F_{q}$. Consider a twisted Reed-Solomon code $\mathcal C_4(\boldsymbol{\alpha}, 1,3,\eta)$, when $\boldsymbol{\alpha}=(1,\gamma^{20}, \gamma^{40}, \gamma^{60},  \gamma,\gamma\gamma^{20}, \gamma\gamma^{40}, \gamma\gamma^{60})$ and $\eta=\gamma^i\in \Bbb F_{81}$. Then its generator matrix $G_1$ is given in the top of next page. 
\begin{figure*}[!h]
\normalsize
\setcounter{mytempeqncnt}{\value{equation}}
\setcounter{equation}{4} 
\begin{equation*}
\resizebox{16cm}{!}{$ G_1=\left(\begin{array}{cccccccccc}
1 &1 &1  &1& 1 &1 &1  &1\\
1&\gamma^{20}& \gamma^{40}&\gamma^{60} & \gamma &\gamma\gamma^{20}& \gamma\gamma^{40}& \gamma\gamma^{60}\\
1&(\gamma^{20})^2& (\gamma^{40})^2&(\gamma^{60})^2 &(\gamma)^2&(\gamma\gamma^{20})^2& (\gamma\gamma^{40})^2& (\gamma\gamma^{60})^2\\
1+\gamma^i&(\gamma^{20})^3+\gamma^i(\gamma^{20})^4& (\gamma^{40})^3+\gamma^i(\gamma^{40})^4&(\gamma^{60})^3+\gamma^i(\gamma^{60})^4 & (\gamma)^3+\gamma^i(\gamma)^4&(\gamma\gamma^{20})^3+\gamma^i(\gamma\gamma^{20})^4& (\gamma\gamma^{40})^3+\gamma^i(\gamma\gamma^{40})^4& (\gamma\gamma^{60})^3+\gamma^i(\gamma\gamma^{60})^4
\end{array}\right)$}
\end{equation*}
\setcounter{equation}{\value{mytempeqncnt}}
\hrulefill 
\vspace*{4pt} 
\end{figure*}
\setcounter{equation}{3}


By Lemma \ref{lem7}, $\mathcal C_4(\boldsymbol{\alpha}, 1,3,\gamma^i)$ is Euclidean LCD for all $i$. By Magma, it follows that
the codes $\mathcal C_4(\boldsymbol{\alpha}, 1,3,\eta)$ are MDS with parameters $[8,4]_{81}$ if and only if $\eta$ belongs to $\{\gamma^j: j=0,1,5,6,7,11,15,16,17,19,20,21,25,26,27,31,35,36,37,\\39,40,41,45,46,47,51,
55,56,57,59,60,61,65,66,67,71,\\75,76,77,79\}$. By Magma the code $\mathcal C_4(\boldsymbol{\alpha}, 1,3,1)$ has  a generator matrix of the form of $[I_4 \mid \mathbf{ A}]$, where 
\begin{eqnarray*}  \mathbf{ A}=\left( \begin{array}{cccccc}
\gamma^7 & \gamma^{32}&\gamma^{56} & \gamma^{78} \\
\gamma^{31} & \gamma^{21}&\gamma^{64} & \gamma^{44} \\
\gamma^{12}& \gamma^{9}&\gamma^{74} & \gamma^{77} \\
\gamma^{60} & \gamma^{49}&\gamma^{52} & \gamma^{79} \\
\end{array} \right).
\end{eqnarray*}
Then it is easy to check that the $3\times 3$ minor of the first three rows and columns
 of $\widetilde{\mathbf{ A}}$ is equal to $\gamma^{26}$, which is confirmed by Magma. By Lemma \ref{lem3}, $\mathcal C_4(\boldsymbol{\alpha}, 1,3,1)$ is an $[8,4]_{81}$ Euclidean LCD MDS non-Reed-Solomon code.
 $\blacksquare$} 

\end{example}

An effective method for construction of twisted Reed-Solomon codes with MDS property
is to use the lifting of the finite field (refer to \cite{BPR}).  We note that the Euclidean LCD property of a given code is preserved under the lifting of the finite field. Hence, we obtain the following theorem.


\begin{theorem}  Let $q$ be a power of an odd prime and $\Bbb F_s \subset \Bbb F_q$.   Let $k$ is a positive integer with $k\mid (q-1)$, $2<k< (s-1)/2$. Let $\boldsymbol{\alpha}=(\alpha_1,\ldots,\alpha_k,\gamma\alpha_1,\ldots,\gamma\alpha_k)$,
where $\gamma$ is a primitive element of $\Bbb F_s$ and $\alpha_i=\gamma^{\frac{s-1}{k}i}$ for $1\le i\le k$. Assume that the minor condition for $\eta\in\Bbb F_q \backslash \Bbb F_s$ of Lemma \ref{lem3} holds.  Then $\mathcal C_k(\boldsymbol{\alpha}, t,h,\eta)$ is a $[2k, k]_q$ Euclidean LCD MDS non-Reed-Solomon code.





\end{theorem}

{\em Proof:} Similar to Lemma \ref{lem7}, by the conditions  the code $\mathcal C_k(\boldsymbol{\alpha}, t,h,\eta)$ is Euclidean LCD. 
By Lemma \ref{lem2} and  $\eta \in\Bbb F_q \backslash \Bbb F_s $, then $\mathcal C_k(\boldsymbol{\alpha}, t,h,\eta)$ is an MDS twisted Reed-Solomon code.  Then the result follows from  Lemma \ref{lem3}.
 $\blacksquare$

In the following, we consider the  construction of Roth and Lempel \cite{RL}.

\begin{lemma} \label{lem8} {\rm  Let $q$ be a power of an odd prime, and let $k\mid (q-1)$ and $k\geq 3$. Let $\gamma$ be a primitive element of $\Bbb F_q$ and $\alpha_i=\gamma^{\frac{q-1}{k}i}$, $1\le i\le k$.
Then there exists a Euclidean LCD Roth-Lempel code $RL({\boldsymbol{ \alpha}},k,n)$
with one of the following parameters:

(1) $[k+2,k]_q$ if $\boldsymbol{\alpha}=(\alpha_1,\ldots,\alpha_k)$;

(2) $[k+3,k]_q$ if  $\gcd(k+1,q)=1$ and $\boldsymbol{\alpha}=(0,\alpha_1,\ldots,\alpha_k)$;

(3) $[2k+2,k]_q$ if $k< \frac{q-1}{2}$ and $\boldsymbol{\alpha}=(\alpha_1,\ldots,\alpha_k,\gamma\alpha_1,\ldots,\gamma\alpha_k)$.
}
\end{lemma}

{\em Proof:}  By  Lemma \ref{lem1}, the  code $RL({\boldsymbol{ \alpha}},k,n)$  over $\Bbb F_q$  is Euclidean LCD if and only if $G_2G_2^T$ is nonsingular.

Let $$D=\left( \begin{array}{cccccc}
 0 &0\\
 0 &0\\
 \vdots &\vdots\\
  0 &0\\
 0 &1\\
 1 &\delta\\
\end{array} \right).$$

(1) Let  $G_2=[A_{1}: D]$, where $A_{1}$ is given in Equation (\ref{eq4}). Then \begin{eqnarray*}G_2G_2^T=A_{1}A_{1}^T+DD^T=\left( \begin{array}{ccccccc}
k & 0& 0&\ldots  & 0 &0\\
0&0&0& \ldots   &  0 &k\\
\vdots&\vdots& \vdots& \ldots  & \vdots &\vdots\\
0 & 0&k&\cdots  &  1 &\delta\\
0& k& 0&\ldots  &\delta &1+\delta^2\\
\end{array} \right).
\end{eqnarray*} Then the matrix $G_2G_2^T$ is nonsingular; so, $RL({\boldsymbol{ \alpha}},k,k+2)$ is a Euclidean LCD code.

(2) Let  $G_2=[e_1:A_{1}: D]$, where $e_1=(1,0,\ldots, 0)^T$. Then \begin{eqnarray*}&&G_2G_2^T=e_1e_1^T+A_{1}A_{1}^T+DD^T\\&=&\left( \begin{array}{ccccccc}
k+1 & 0& 0&\ldots  & 0 &0\\
0&0&0& \ldots   &  0 &k\\
\vdots&\vdots& \vdots& \ldots  & \vdots &\vdots\\
0 & 0&k&\cdots  &  1 &\delta\\
0& k& 0&\ldots  &\delta &1+\delta^2\\
\end{array} \right).
\end{eqnarray*} Then the matrix $G_2G_2^T$ is nonsingular; hence,  $RL({\boldsymbol{ \alpha}},k,k+3)$ is a Euclidean LCD code.

(3) Note that $k\neq \frac{q-1}{2}$. Let $G_2=[A_{1}: A_{\gamma}: D]$.
Then \begin{eqnarray*}&&G_2G_2^T=A_{1}A_{1}^T+A_{\gamma}A_{\gamma}^T+DD^T\\
&=&\left( \begin{array}{ccccccc}
2k & 0& 0&\ldots  & 0 &0\\
0&0&0& \ldots   &  0 &(1+\gamma^k) k\\
\vdots&\vdots& \vdots& \ldots  & \vdots &\vdots\\
0 & 0&(1+\gamma^k) k&\cdots  &  1 &\delta\\
0& (1+\gamma^k) k& 0&\ldots  &\delta &1+\delta^2\\
\end{array} \right).
\end{eqnarray*}
 By the proof of Lemma \ref{lem7},  the matrix $G_2G_2^T$ is nonsingular; therefore, $RL({\boldsymbol{ \alpha}},k,2k+2)$ is a Euclidean  LCD code. This completes the proof. 
 $\blacksquare$

\begin{theorem} {\rm  Let $q$ be a power of an odd prime and  $\Bbb F_s \subset \Bbb F_q$.
Let $k\geq3$ be an integer with $k\mid (s-1)$. 

(1) If  $\gcd(k+1,s)=1$, then there exists a $[k+3, k]_q$ Euclidean LCD MDS non-Reed-Solomon code.

(2) If $k< \frac{s-1}{2}$, then there exists a $[2k+2, k]_q$ Euclidean LCD MDS non-Reed-Solomon code.}
\end{theorem}

{\em Proof:} By the proof of Lemma \ref{lem8}, we can construct a Euclidean LCD Roth-Lempel  code  over $\Bbb F_q$. Note that we can require that $\alpha_i\in \Bbb F_s$ for all $i$. By Lemma 2.12, an Roth-Lempel code is a non-Reed-Solomon code, and it is an MDS code if and only if the set $S=\{\alpha_1, \ldots, \alpha_n\}$ forms an $(n,k-1,\delta)$-set in $\Bbb F_q$; that is, there exists an element $\delta\in \Bbb F_q$ such that no $k-1$ elements of $S$ sum to $\delta$. Note that we can require that $\alpha_i\in \Bbb F_s$ for all $i$. Hence, $S \subseteq \Bbb F_s$, and we can find some $\delta\in \Bbb F_q \backslash \Bbb F_s$ such that $S$ is an $(n,k-1,\delta)$-set in $\Bbb F_q$.
Similar to Lemma \ref{lem8}, by the conditions we can find a vector $\boldsymbol{\alpha}\in \Bbb F_s^n$ such that the code $\mathcal C_k(\boldsymbol{\alpha}, t,h,\eta)$ is Euclidean LCD and the result follows.
 $\blacksquare$

The followings are examples of  Theorem 3.6.
\begin{example}{\rm (1) Let $q=3^2=9$ and $k=4$. Let $\gamma$ be a primitive element of $\Bbb F_{9}$.
We choose $\boldsymbol{\alpha}=(0,1, \gamma^{2}, \gamma^{4},\gamma^{6} )$ and $\delta=\gamma^i$ for some integer $i$ with $0\le i\le 7$. Then the generator matrix  of the code $\mathcal{C}_2$  is given as follows:
\begin{eqnarray*} \left(\begin{array}{ccccccccc}
1 &1 &1  &1&1&0&0\\
0 &1 &\gamma^{2} &\gamma^{4}&\gamma^{6}&0&0\\
0 &1 &(\gamma^{2})^2 &(\gamma^{4})^2&(\gamma^{6})^2&0&1\\
0 &1 &(\gamma^{2})^3 &(\gamma^{4})^3&(\gamma^{6})^3&1&\gamma^i\\
\end{array}\right)
\end{eqnarray*} By Magma, there is no $i$ such  that
$\mathcal{C}_2$ is a Euclidean LCD MDS non-Reed-Solomon Roth-Lempel code with parameters $[7,4]_9$. However, we can make these LCD codes have the MDS property by lifting the finite field $\Bbb F_{9}$ which is shown in the following.

 (2) Let $q=3^4$ and $k=4$. Let $w$ be a primitive element of $\Bbb F_{81}$ and  $\gamma$ a primitive element of $\Bbb F_{9}$ with $\gamma=w^{10}$. Choose $\boldsymbol{\alpha}=(0,1, \gamma^{2}, \gamma^{4},\gamma^{6} )$  and $\delta=w^i\in \Bbb F_{81}$.
 Then the generator matrix of $RL({\boldsymbol{ \alpha}},k,n)$ is given as follows:
\begin{eqnarray*} \left(\begin{array}{ccccccccc}
1 &1 &1  &1&1&0&0\\
0 &1 &\gamma^{2} &\gamma^{4}&\gamma^{6}&0&0\\
0 &1 &(\gamma^{2})^2 &(\gamma^{4})^2&(\gamma^{6})^2&0&1\\
0 &1 &(\gamma^{2})^3 &(\gamma^{4})^3&(\gamma^{6})^3&1&w^i\\
\end{array}\right).
\end{eqnarray*}
By Theorem 3.6 $(1)$, $RL({\boldsymbol{ \alpha}},k,k+3)$ is a Euclidean LCD MDS non-Reed-Solomon code with parameters $[7,4]_{81}$ when $i$ is not divisible by 10.
}


\end{example}


\begin{remark}
{\rm
We emphasize that any Euclidean LCD MDS code of non-Reed-Solomon type constructed in Theorems 3.4 and 3.6 is not monomially equivalent to any Euclidean  LCD code constructed by the method of Carlet et al.~\cite{CMTQP}.

Now, we briefly justify why they are not monomially equivalent for the case of Theorem 3.4, and the case of Theorem 3.6 can be also justified similarly.
According to the result of Carlet et al. ~\cite[Theorem 5.1]{CMTQP}, assume that there is a $[2k,k]$ linear MDS code over $\mathbb{F}_q$ with generator matrix $[I_k~ A]$ satisfying the conditions of~\cite[Theorem 5.1]{CMTQP}.
Then there is a monomial matrix $M$ such that $[I_k~ A]M$ generates a LCD MDS code $\mathcal C$ by~\cite[Theorem 5.1]{CMTQP}. 
Now, if we suppose that the code $\mathcal C$ is monomially equivalent to our LCD MDS code $\mathcal{C}_k(\boldsymbol{\alpha},t,h,\eta)$ of non-Reed-Solomon type with generator matrix $G_1$, then there should exist a monomial matrix $M'$ such that $G_1=[I_k~ A]MM'$ or $G_1(M')^{-1}M^{-1}=[I_k~A]$. Recall that a monomial matrix is a square matrix which has exactly one nonzero entry in each row and each column. It follows that $MM'=PDP'$, where $P$ and $P'$ are permutation matrices and $D$ is a diagonal matrix. Therefore, the entries of the first row of $G_1(M')^{-1}M^{-1}$ are nonzero and all-one except two coordinate positions, and the product of all the entries of the first row of $[I_k~A]$ is zero; this is impossible. }  

\end{remark}

\subsection{Hermitian LCD MDS codes} 

In this subsection, we consider Hermitian LCD MDS codes over $\Bbb F_{q^2}$.

Let $\gamma$ be a primitive element of $\Bbb F_{q^2}$ and $k\mid (q^2-1)$. Then $\gamma^{\frac{q^2-1}{k}}$ generates a subgroup of order $k$ in $\Bbb F_{q^2}^*$. Let $\alpha_i=\gamma^{\frac{q^2-1}{k}i}$ with $1\le i\le k$.

For $i,j\in \{0,\ldots, k-1\}$, assume that $a_{\beta}(i,j)$ is the entry in the $(i+1)$-th row and $(j+1)$-th column of the matrix $A_{\beta}\overline{A}_{\beta}^T$, where $A_{\beta}$ is given in Equation (\ref{eq4}). Then
\begin{eqnarray*} &&a_{\beta}(i,j)=(\beta\alpha_1)^i(\overline{\beta\alpha_1})^j+\cdots+(\beta\alpha_k)^i(\overline{\beta\alpha_k})^j\\
&=&\left\{
\begin{array}{ll}
  \beta^{i+jq}k   &      \mbox{if}\ i+jq\equiv0\pmod {k},\\
0 & \mbox{otherwise}.
\end{array} \right. \end{eqnarray*}
Every row of the matrix $A_{\beta}\overline{A}_{\beta}^T$ has exactly one nonzero element, and every column of the matrix $A_{\beta}\overline{A}_{\beta}^T$ has exactly one nonzero element. Hence, the matrix $A_{\beta}\overline{A}_{\beta}^T$ is nonsingular over $\Bbb F_{q^2}$.


The following lemma plays an important role in proving our main results of this Subsection 3.2.
First, we investigate the  twisted Reed-Solomon code $\mathcal C_k(\boldsymbol{\alpha}, t,h,\eta)$ over $\Bbb F_{q^2}$.

\begin{lemma}{\rm Let $q$ be a power of an odd prime and $k$ be a positive integer with $k\mid (q^2-1)$. If there exists an odd prime number $p$ such that $v_p(k)<v_p(q^2-1)$ and $h>0$, then  there exists a $[2k,k]_{q^2}$ Hermitian  LCD twisted Reed-Solomon code $\mathcal C_k(\boldsymbol{\alpha}, t,h,\eta)$ over $\Bbb F_{q^2}$ for
$\boldsymbol{\alpha}=(\alpha_1,\ldots,\alpha_k,\gamma^r\alpha_1,\ldots,\gamma^r\alpha_k)$,
where $\gamma$ is a primitive element of $\Bbb F_{q^2}$, $\alpha_i=\gamma^{\frac{q^2-1}{k}i}$, $1\le i\le k$, and $r=2^{v_2(q^2-1)}$.

}
\end{lemma}

{\em Proof:} The generator matrix $G_1$ of the twisted Reed-Solomon code $\mathcal C_k(\boldsymbol{\alpha}, t,h,\eta)$ over $\Bbb F_{q^2}$ is shown in Equation (\ref{eq1}).  By  Lemma \ref{lem1},  $\mathcal C_k(\boldsymbol{\alpha}, t,h,\eta)$   is Hermitian LCD if and only if $G_1\overline{G}_1^T$ is nonsingular.  Let $E=\theta_{h+hq}+\eta^q\theta_{h+lq}+\eta\theta_{l+hq}+\eta^{1+q}\theta_{l+lq}, \theta_j=\sum_{i=1}^{n}\alpha_i^j$ and $l=k-1+t$. Then we compute $G_1\overline{G}_1^T$ in Equation (9) in the top of next page.
\begin{figure*}[!h]
\normalsize
\setcounter{mytempeqncnt}{\value{equation}}
\setcounter{equation}{8} 
\begin{equation}
\label{eq9}
\resizebox{16cm}{!}{$G_1\overline{G}_1^T=\left( \begin{array}{cccccccccccccc}
n & \theta_{q}& \ldots  &\theta_{(h-1)q}&\theta_{hq}+\eta^q\theta_{lq} & \theta_{(h+1)q}  &\ldots& \theta_{(k-2)q} &\theta_{(k-1)q}\\
\theta_{1} & \theta_{1+q}& \ldots  &\theta_{1+(h-1)q}&\theta_{1+hq}+\eta^q\theta_{1+lq} & \theta_{1+(h+1)q}  &\ldots& \theta_{1+(k-2)q} &\theta_{1+(k-1)q}\\
\vdots & \vdots& \ldots  &\vdots&\vdots &  \vdots& \ldots  &\vdots&\vdots\\
\theta_{h-1} & \theta_{h-1+q}& \ldots  &\theta_{h-1+(h-1)q}&\theta_{h-1+hq}+\eta\theta_{h-1+lq} & \theta_{h-1+(h+1)q}  &\ldots& \theta_{h-1+(k-2)q} &\theta_{h+(k-1)q}\\
\theta_{h} +\eta\theta_{l}& \theta_{h+q} +\eta\theta_{l+q}& \ldots  &\theta_{h+(h-1)q} +\eta\theta_{h-1+lq}&E & \theta_{h+(h+1)q}+\eta\theta_{l+(h+1)q}& \ldots& \theta_{h+(k-2)q}+\eta\theta_{k-2+lq}&\theta_{h+(k-1)q}+\eta\theta_{l+(k-1)q}\\
\theta_{h+1} & \theta_{h+1+q}& \ldots  &\theta_{h+1+(h-1)q}&\theta_{h+1+hq}+\eta\theta_{h+1+lq} & \theta_{h+1+(h+1)q}  &\ldots& \theta_{h+1+(k-2)q} &\theta_{h+1+(k-1)q}\\
\vdots & \vdots& \ldots  &\vdots&\vdots &  \vdots& \ldots  &\vdots&\vdots\\
\theta_{k-1} & \theta_{k-1+q}& \ldots  &\theta_{k-1+(h-1)q}&\theta_{k-1+hq}+\eta\theta_{k-1+lq} &
\theta_{k-1+(h-1)q}  &\ldots&\theta_{k-1+(k-2)q} &\theta_{k-1+(k-1)q}\\
\end{array} \right),$}
\end{equation}
\setcounter{equation}{\value{mytempeqncnt}}
\hrulefill 
\vspace*{4pt} 
\end{figure*}
\setcounter{equation}{3}


Let $C_{\beta}=A_{\beta}+B_{\beta}$, where $A_{\beta}$ is given in Equation (\ref{eq4}) and $B_{\beta}$ is given in Equation (\ref{eq7}).  By Equation (\ref{eq9}), we compute $C_{\beta}\overline{C}_{\beta}^T$ in Equation (10) in the top of next page.
\begin{figure*}[!h]
\normalsize
\setcounter{mytempeqncnt}{\value{equation}}
\setcounter{equation}{9} 
\begin{equation}
\label{eq10}
\resizebox{16cm}{!}{$C_{\beta}\overline{C}_{\beta}^T=A_{\beta}\overline{A}_{\beta}^T+\left( \begin{array}{ccccccccc}
0 & 0& \ldots  &0 & \eta^q\beta^{lq}\theta_{lq}&0  &\ldots&0\\
0 & 0& \ldots  &0 & \eta\beta^{1+lq}\theta_{1+lq}&0 &\ldots&0\\
\vdots & \vdots& \ldots  &\vdots & \vdots & \vdots& \ldots  &\vdots\\
0 & 0& \ldots  &0 & \eta\beta^{h-1+lq}\theta_{h-1+lq}&0  &\ldots& 0\\
\eta\beta^l\theta_{l}& \eta\beta^{l+q}\theta_{l+q}& \ldots  &\eta\beta^{l+(h-1)q}\theta_{l+(h-1)q} & \eta\beta^{l+hq}\theta_{l+hq}+\eta^{1+q}\beta^{l+lq}\theta_{l+lq}+\eta^q\beta^{h+lq}\theta_{h+lq}&\eta\beta^{l+(h+1)q}\theta_{l+(h+1)q}&\ldots&\eta\beta^{l+(k-1)q}\theta_{l+(k-1)q}\\

0 &0& \ldots  &0 & \eta\beta^{h+1+lq}\theta_{h+1+lq}&0 &\ldots&0\\
\vdots & \vdots& \ldots  &\vdots & \vdots & \vdots& \ldots  &\vdots\\
0 & 0& \ldots  &0 & \eta\beta^{k-1+lq}\theta_{k-1+lq}&0 &\ldots&0

\end{array} \right).$}
\end{equation}
\setcounter{equation}{\value{mytempeqncnt}}
\hrulefill 
\vspace*{4pt} 
\end{figure*}
\setcounter{equation}{3}



 Note that $\alpha _i^{q^2}=\alpha_i$. Then $\overline{\{\theta_l, \theta_{l+q},\ldots, \theta_{l+(k-1)q}\}}=\{\theta_{lq}, \theta_{1+lq},\ldots, \theta_{(k-1)+lq}\}$.  Exactly one element in the set $\{\theta_l, \theta_{l+q},\ldots, \theta_{l+(k-1)q}\}$ has value $k$. Hence,  \begin{eqnarray*}C_{\beta}\overline{C}_{\beta}^T=A_{\beta}\overline{A}_{\beta}^T+\left( \begin{array}{cccccccc}
0 &\ldots&0& \ldots  & 0 &\ldots&0\\
\vdots&& \vdots& \ldots  & \vdots &&\vdots\\
0&\ldots &0& \ldots   & \overline{*_{\beta} } & \ldots&0 \\
\vdots&& \vdots& \ldots  & \vdots &&\vdots\\
0&\ldots &*_{\beta}& \ldots   & \Delta_{\beta}  &\ldots& 0 \\
\vdots&& \vdots& \ldots  & \vdots &&\vdots\\
0&\ldots&0&\cdots  & 0&\ldots&0 \\
\end{array} \right),\end{eqnarray*} where $\ast_{\beta}$ and $\Delta_{\beta}$ are elements belong to $\Bbb F_{q^2}$, $\ast_{\beta}$, $ \overline{*_{\beta} }$, and $\Delta_{\beta}$ are entries placed in the $(i+1,h+1)th$, $(h+1,i+1)$th, and $(h+1,h+1)$th positions, respectively and the other elements are all zero.





Let $h>0$ and $G_1=[C_1:  C_{\gamma^r}]$, where $r=2^{v_2(q^2-1)}$. By the condition that there exists an odd prime number $p$ such that $v_p(k)<v_p(q^2-1)$, any two columns in $G_1$ are not same.
Then \begin{eqnarray*}&&G_1\overline{G}_1^T=C_1\overline{C}_1^T+C_{\gamma^r}\overline{C}_{\gamma^r}^T\\
&=&A_{1}\overline{A}_{1}^T+A_{\gamma^r}\overline{A}_{\gamma^r}^T\\
&+&\left( \begin{array}{cccccccc}
0 &\ldots&0& \ldots  & 0 &\ldots&0\\
\vdots&& \vdots& \ldots  & \vdots &&\vdots\\
0&\ldots &0& \ldots   & \overline{*_{1}} + \overline{*_{\gamma^r}}& \ldots&0 \\
\vdots&& \vdots& \ldots  & \vdots &&\vdots\\
0&\ldots &*_{1} + *_{\gamma^r}& \ldots   &  \Delta_1+\Delta_{\gamma^r}  &\ldots& 0 \\
\vdots&& \vdots& \ldots  & \vdots &&\vdots\\
0&\ldots&0&\cdots  & 0&\ldots&0 \\
\end{array} \right).\end{eqnarray*}

Let $b(i,j)$ be the entry in the $i$-th row and $j$-th column of the matrix $A_{1}\overline{A}_{1}^T+A_{\gamma^r}\overline{A}_{\gamma^r}^T$. Then \begin{equation*} b(i,j)=\left\{
\begin{array}{ll}
 (1+ \gamma^{r(i+jq)})k   &      \mbox{if}\ i+jq\equiv0\pmod {k},\\
0 & \mbox{otherwise}.
\end{array} \right. \end{equation*}
When $i+jq\equiv0\pmod {k}$, assume that $b(i,j)=0$. We have $\gamma^{r(i+jq)}=-1=\gamma^{\frac{q^2-1}{2}}$ and $r(i+jq)\equiv \frac{q^2-1}{2}\pmod{q^2-1}$. Since $r=2^{v_2(q^2-1)}$ and $v_2({\frac{q^2-1}{2}})=v_2(q^2-1)-1$, we get a contradiction. So, we have $b(i,j)\neq0$ when $i+jq\equiv0\pmod {k}$.
Therefore, every row of the matrix $A_{1}\overline{A}_{1}^T+A_{\gamma^r}\overline{A}_{\gamma^r}^T$ has a nonzero element and every column of the matrix $A_{1}\overline{A}_{1}^T+A_{\gamma^r}\overline{A}_{\gamma^r}^T$ has exactly one nonzero element.

Since the matrix $G_1\overline{G}_1^T$ is  conjugate symmetric,
we can delete the elements $*_{1} + *_{\gamma^r}$ and $\overline{*_{1}} + \overline{*_{\gamma^r}} $ by some elementary row and column operations of matrices at the same time. Therefore, we can find an elementary matrix $P$ such that \begin{eqnarray*}&&PG_1\overline{G}_1^T\overline{P}^T=PA_{1}\overline{A}_{1}^T\overline{P}^T+PA_{\gamma^r}\overline{A}_{\gamma^l}^T\overline{P}^T\\
&+&\left( \begin{array}{cccccccc}
0 &\ldots&0& \ldots  & 0 &\ldots&0\\
\vdots&& \vdots& \ldots  & \vdots &&\vdots\\
0&\ldots &0& \ldots   & 0& \ldots&0 \\
\vdots&& \vdots& \ldots  & \vdots &&\vdots\\
0&\ldots &0& \ldots   &  \Delta_1+\Delta_{\gamma^r}  &\ldots& 0 \\
\vdots&& \vdots& \ldots  & \vdots &&\vdots\\
0&\ldots&0&\cdots  & 0&\ldots&0 \\
\end{array} \right).\end{eqnarray*}

 Hence, $G_1\overline{G}_1^T$ is nonsingular; thus, the code $\mathcal C_k(\boldsymbol{\alpha}, t,h,\eta)$ over $\Bbb F_{q^2}$ is a Hermitian LCD code. This completes the proof.
 $\blacksquare$





\begin{example}{\rm Let $q=11^2=121$,  $k=5$, and $\gamma$ be a primitive element of the finite field $\Bbb F_{121}$.  Consider a
twisted Reed-Solomon code $\mathcal C_5(\boldsymbol{\alpha}, 1,3,\eta)$ with $\boldsymbol{\alpha}=(1,\gamma^{24}, \gamma^{48}, \gamma^{72}, \gamma^{96} ,\gamma^8,\gamma^{8}\gamma^{24}, \gamma^8\gamma^{48}, \gamma^8\gamma^{72}, \gamma^8\gamma^{96})$ and $\eta=\gamma^i\in \Bbb F_{121}$.Then its generator matrix  $G_2$ is given in the top of next page. By Lemma 3.8, $\mathcal C_5(\boldsymbol{\alpha}, 1,3,\gamma^i)$ is Hermitian LCD for all $i$.
  By Magma,
the codes $\mathcal C_5(\boldsymbol{\alpha}, 1,3,\gamma^i)$ are MDS with parameters $[10,5]_{121}$ when $\eta\in \{\gamma^j: j=0,5,7,10,13,22,23\}$. By Magma the code $\mathcal C_5(\boldsymbol{\alpha}, 1,3,\gamma^{23})$ has a generator matrix of the form of $[I_5 \mid \mathbf{ A}]$, where 
\begin{eqnarray*}  \mathbf{ A}=\left( \begin{array}{cccccc}
\gamma^{111} & \gamma^{115}&\gamma^{6} & \gamma^{45}&10 \\
\gamma^{73} & \gamma^{19}&\gamma^{5} & \gamma^{54}&\gamma^{81} \\
\gamma^{91} & \gamma^{10}&\gamma^{22} & \gamma^{55}&\gamma^{81} \\
\gamma^{40} & \gamma^{94}&\gamma^{7} & \gamma^{43}&\gamma^{62} \\
\gamma^{38} & \gamma^{38}&\gamma^{116} & \gamma^{104}&\gamma^{56} \\
\end{array} \right).
\end{eqnarray*}
Then it is easy to check that the $3\times 3$ minor of the first three rows and columns
 of $\widetilde{\mathbf{ A}}$ is equal to $8$, which is confirmed by Magma. By Lemma \ref{lem3}, $\mathcal C_5(\boldsymbol{\alpha}, 1,3,\gamma^{23})$ is a $[10,5]_{121}$ Hermitian LCD MDS non-Reed-Solomon code.
$\blacksquare$

\begin{figure*}[!h]
\normalsize
\setcounter{mytempeqncnt}{\value{equation}}
\setcounter{equation}{6} 
\begin{equation*}
\label{eq10}
\resizebox{18cm}{!}{$G_2=\left(\begin{array}{cccccccccccccc}
1 &1 &1  &1&1  & 1 &1 &1  &1&1\\
1&\gamma^{24}&\gamma^{48}&\gamma^{72}&\gamma^{96} & \gamma^8&\gamma^8\gamma^{24}&\gamma^8\gamma^{48}&\gamma^8\gamma^{72}&\gamma^8\gamma^{96}\\
1&(\gamma^{24})^2&(\gamma^{48})^2&(\gamma^{72})^2&(\gamma^{96})^2 & (\gamma^8)^2&(\gamma^8\gamma^{24})^2&(\gamma^8\gamma^{48})^2&(\gamma^8\gamma^{72})^2&(\gamma^8\gamma^{96})^2\\
1+\gamma^i&(\gamma^{24})^3+\gamma^i(\gamma^{24})^6&(\gamma^{48})^3+\gamma^i(\gamma^{48})^6&(\gamma^{72})^3+\gamma^i(\gamma^{72})^6&(\gamma^{96})^3+\gamma^i(\gamma^{96})^6 & (\gamma^8)^3+\gamma^i(\gamma^8)^6&(\gamma^8\gamma^{24})^3+\gamma^i(\gamma^8\gamma^{24})^6&(\gamma^8\gamma^{48})^3+\gamma^i(\gamma^8\gamma^{48})^6&(\gamma^8\gamma^{72})^3+\gamma^i(\gamma^8\gamma^{72})^6&(\gamma^8\gamma^{96})^3+\gamma^i(\gamma^8\gamma^{96})^6\\
1&(\gamma^{24})^4&(\gamma^{48})^4&(\gamma^{72})^4&(\gamma^{96})^4 & (\gamma^8)^4&(\gamma^8\gamma^{24})^4&(\gamma^8\gamma^{48})^4&(\gamma^8\gamma^{72})^4&(\gamma^8\gamma^{96})^4\\
\end{array}\right).$}
\end{equation*}
\setcounter{equation}{\value{mytempeqncnt}}
\hrulefill 
\vspace*{4pt} 
\end{figure*}
\setcounter{equation}{3}

}
\end{example}

In a similar way as Theorem 3.4, we obtain the following result.

\begin{theorem}  {\rm    Let $q$ be a power of an odd prime and  $\Bbb F_s \subset \Bbb F_{q^2}$. Let $k$ be a positive integer such that $k\mid (s-1)$,  $2<k< (s-1)/2$. There exists an odd prime number $p$ such that $v_p(k)<v_p(s-1)$. Let 
$\boldsymbol{\alpha}=(\alpha_1,\ldots,\alpha_k,\gamma^r\alpha_1,\ldots,\gamma^r\alpha_k)$,
where $\gamma$ is a primitive element of $\Bbb F_{s}$, $\alpha_i=\gamma^{\frac{s-1}{k}i}$, $1\le i\le k$, and $r=2^{v_2(s-1)}$. Assume that the minor condition for $\eta\in\Bbb F_{q^2} \backslash \Bbb F_s$ of Lemma \ref{lem3} holds.  Then $\mathcal C_k(\boldsymbol{\alpha}, t,h,\eta)$ is a $[2k, k]_{q^2}$ Hermitian LCD MDS non-Reed-Solomon code.




}
\end{theorem}

{\em Proof:} By Lemma 3.9, $\mathcal C_k(\boldsymbol{\alpha}, t,h,\eta)$ over $\Bbb F_{q^2}$ is Hermitian LCD.  By Lemma \ref{lem2} and $\eta \in\Bbb F_{q^2} \backslash \Bbb F_s $, then the twisted Reed-Solomon code $\mathcal C_k(\boldsymbol{\alpha}, t,h,\eta)$ over $\Bbb F_{q^2}$ is MDS. Then the result follows from  Lemma \ref{lem3}. $\blacksquare$


In the following, we consider Roth-Lempel codes.

\begin{lemma} {\rm  Let $q$ be a power of an odd prime, and let $k\mid (q^2-1)$ and $k\geq 3$. Let $\gamma$ be a primitive element of $\Bbb F_{q^2}$ and $\alpha_i=\gamma^{\frac{q^2-1}{k}i}$ for $1\le i\le k$. Then there exists a Hermitian LCD Roth-Lempel code $RL({\boldsymbol{ \alpha}},k,n)$  over $\Bbb F_{q^2}$   with one of the following parameters:

(1) $[k+2,k]_{q^2}$ if $\boldsymbol{\alpha}=(\alpha_1,\ldots,\alpha_k)$;

(2) $[k+3,k]_{q^2}$ if  $\gcd(k+1,q)=1$ and $\boldsymbol{\alpha}=(0,\alpha_1,\ldots,\alpha_k)$;

(3) $[2k+2,k]_{q^2}$ if there exists an odd prime number $p$ such that $v_p(k)<v_p(q^2-1)$, $\boldsymbol{\alpha}=(\alpha_1,\ldots,\alpha_k,\gamma^r\alpha_1,\ldots,\gamma^r\alpha_k)$, and $r=2^{v_2(q^2-1)}$.






}

\end{lemma}

{\em Proof:} By  Lemma \ref{lem1}, the Roth-Lempel code  over $\Bbb F_q$ in Definition 2.11 is Hermitian LCD if and only if $G_2\overline{G}_2^T$ is nonsingular.

(1) Let  $G_2=[A_{1}: D]$, where  $D$ is given in the proof of Lemma 3.5. Then \begin{eqnarray*}&&G_2\overline{G}_2^T=A_{1}\overline{A}_{1}^T+D\overline{D}^T\\
&=&A_{1}\overline{A}_{1}^T+\left( \begin{array}{ccccccc}
0 & 0& 0&\ldots  & 0 &0\\
\vdots&\vdots& \vdots& \ldots  & \vdots &\vdots\\
0 & 0&0&\cdots  &  1 &\delta^q\\
0& 0& 0&\ldots  &\delta &1+\delta^{1+q}\\
\end{array} \right).
\end{eqnarray*} Since the matrix $A_{1}\overline{A}_{1}^T$ is nonsingular,  the matrix $G_2\overline{G}_2^T$ is nonsingular and   $RL({\boldsymbol{ \alpha}},k,k+2)$ is a Hermitian LCD code.

(2) Let  $G_2=[e_1:A_{1}: D]$, where $e_1=(1,0,\ldots, 0)^T$. Then $G_2\overline{G}_2^T=e_1e_1^T+A_{1}\overline{A}_{1}^T+D\overline{D}^T$ and the entry of the $(1,1)$th position of $A_{1}\overline{A}_{1}^T$ is $k$, which is nonzero. Then the entry of the $(1,1)$th position of the matrix $G_2\overline{G}_2^T$ is  $k+1$, and so the matrix $G_2\overline{G}_2^T$ is nonsingular. Therefore, the $RL({\boldsymbol{ \alpha}},k,k+3)$ code is a Hermitian LCD code.

(3)  By the proof of Lemma 3.5, let  $G_2=[A_1:  A_{\gamma^r}: D]$.
By the proof of Lemma 3.1,  we can take some element $\delta\in \Bbb F_{q^2}$ such that the matrix $G_2\overline{G}_2^T$ is nonsingular, and  hence the $RL({\boldsymbol{ \alpha}},k,2k+2)$ code is a Hermitian  LCD code. 

This completes the proof.
 $\blacksquare$

In a similar way as Theorem 3.6, we have the following theorem on Hermitian LCD codes over $\Bbb F_{q^2}$.

\begin{theorem}{\rm Let $q$ be a power of an odd prime and  $ \Bbb F_{q^2}$ be the finite field of order $q^2$. Let $k$ be an integer such that $k\mid (q-1)$ and $k\ge 3$.

(1) If  $\gcd(k+1,q)=1$, then there exists a $[k+3, k]_{q^2}$ Hermitian LCD MDS non-Reed-Solomon code over $\Bbb F_{q^2}$.

(2)  If there exists an odd prime number $p$ such that $v_p(k)<v_p(q-1)$, then there exists a $[2k+2, k]_{q^2}$ Hermitian LCD MDS non-Reed-Solomon code over $\Bbb F_{q^2}$.


}
\end{theorem}

We give the following example.
\begin{example}{\rm
(1) Let $q=5^2$ and $k=6$. Let $\gamma$ be a primitive element of the finite field $\Bbb F_{25}$, $\boldsymbol{\alpha}=(0,1,\gamma^{4}, \gamma^{8}, \gamma^{12},\gamma^{16}, \gamma^{20} )$,  and $\delta=\gamma^i$ for some integer $i$ with  $0\le i\le 23$. Then the generator matrix of the code $\mathcal{C}_1$  is given as follows:
\begin{eqnarray*}\resizebox{8.3cm}{!}{ $ \left(\begin{array}{cccccccccc}
1 &1 &1  &1&1&1&1&0&0\\
0 &1&\gamma^{4} &\gamma^{8} &\gamma^{12}&\gamma^{16}&\gamma^{20}&0&0\\
0 &1&(\gamma^{4})^2 &(\gamma^{8})^2 &(\gamma^{12})^2&(\gamma^{16})^2&(\gamma^{20})^2&0&0\\
0 &1&(\gamma^{4})^3 &(\gamma^{8})^3 &(\gamma^{12})^3&(\gamma^{16})^3&(\gamma^{20})^3&0&0\\
0 &1&(\gamma^{4})^4 &(\gamma^{8})^4 &(\gamma^{12})^4&(\gamma^{16})^4&(\gamma^{20})^4&0&1\\
0 &1&(\gamma^{4})^5 &(\gamma^{8})^5 &(\gamma^{12})^5&(\gamma^{16})^5&(\gamma^{20})^5&1&\gamma^i\\
\end{array}\right).$}
\end{eqnarray*} By Magma, $\mathcal{C}_1$ is a Hermitian LCD MDS non-Reed-Solomon code over $\Bbb F_{25}$ with parameters $[9,6]_{25}$ when $i=1,2,5,6,9,10,13,14,17,18,21,22$.

(2) Let $q=7^2$ and $k=8$. Let $\gamma$ be a primitive element of the finite field $\Bbb F_{25}$ and $\boldsymbol{\alpha}=(0,1,\gamma^{6}, \gamma^{12}, \gamma^{18},\gamma^{24}, \gamma^{30},\gamma^{36},\gamma^{42} )$,  and $\delta=\gamma^i$ for some integer $i$ with  $0\le i\le 47$. Then the generator matrix of the code $\mathcal{C}_2$  is given as follows:
\begin{equation*}\resizebox{8.3cm}{!}{ $\left(\begin{array}{cccccccccccc}
1 &1 &1  &1&1&1&1&1&1&0&0\\
0 &1&\gamma^{6} &\gamma^{12} &\gamma^{18}&\gamma^{24}&\gamma^{30}&\gamma^{36}&\gamma^{42}&0&0\\
0 &1&(\gamma^{6})^2 &(\gamma^{12} )^2&(\gamma^{18})^2&(\gamma^{24})^2&(\gamma^{30})^2&(\gamma^{36})^2&(\gamma^{42})^2&0&0\\
0 &1&(\gamma^{6})^3 &(\gamma^{12} )^3&(\gamma^{18})^3&(\gamma^{24})^3&(\gamma^{30})^3&(\gamma^{36})^3&(\gamma^{42})^3&0&0\\
0 &1&(\gamma^{6})^4 &(\gamma^{12} )^4&(\gamma^{18})^4&(\gamma^{24})^4&(\gamma^{30})^4&(\gamma^{36})^4&(\gamma^{42})^4&0&0\\
0 &1&(\gamma^{6})^5 &(\gamma^{12} )^5&(\gamma^{18})^5&(\gamma^{24})^5&(\gamma^{30})^5&(\gamma^{36})^5&(\gamma^{42})^5&0&0\\
0 &1&(\gamma^{6})^6 &(\gamma^{12} )^6&(\gamma^{18})^6&(\gamma^{24})^6&(\gamma^{30})^6&(\gamma^{36})^6&(\gamma^{42})^6&0&1\\
0 &1&(\gamma^{6})^7 &(\gamma^{12} )^7&(\gamma^{18})^7&(\gamma^{24})^7&(\gamma^{30})^7&(\gamma^{36})^7&(\gamma^{42})^7&1&\gamma^i\\
\end{array}\right)$  }. \end{equation*}
 By Magma, $\mathcal{C}_2$ is a Hermitian LCD MDS non-Reed-Solomon code over $\Bbb F_{49}$  with parameters $[11,8]_{49}$ when $i=4,16,22,28,29,34,40,46$.  $\blacksquare$



}
\end{example}

\begin{remark}
{\rm We point out that any {\it Hermitian} LCD MDS code of non-Reed-Solomon type constructed in Theorems 3.11 and 3.13 is not monomially equivalent to any {\it Hermitian} LCD code constructed by the method of Carlet et al.~\cite{CMTQP}.
This can be justified in a similar way as Remark 3.8 (for the Euclidean case).}

\end{remark}

\section{Concluding remarks}
Main contributions of this paper are constructions of some new  Euclidean and Hermitian LCD MDS codes of non-Reed-Solomon type.
According to the results of Carlet et al. ~\cite{CMTQP}, all parameters of Euclidean LCD codes $(q > 3)$ and Hermitian LCD codes $(q > 2)$ have been completely determined, including the LCD MDS codes.
However, in the coding theory, it is an important issue to find all {\it inequivalent} codes of the same parameters.
We emphasize that any Euclidean (or Hermitian) LCD MDS code of non-Reed-Solomon type constructed by our method is not monomially equivalent to any Euclidean (or Hermitian) LCD code constructed by the method of Carlet et al. in \cite{CMTQP}; this is justified in Remarks 3.8 and 3.15. 
Finally, we provided some examples of non-Reed-Solomon LCD MDS codes.

 \bigskip
\section*{Acknowledgments}
The authors are very grateful to the reviewers and the
Associate Editor for their valuable comments and suggestions
to improve the quality of this paper.

\ifCLASSOPTIONcaptionsoff
  \newpage
\fi



%

%
\begin{IEEEbiographynophoto}{Yansheng Wu} received his Ph.D. degree from the Nanjing University of Aeronautics and Astronautics, Nanjing, China, in 2019.  From September  2019 to  August 2020, he was a Post-Doctoral Researcher with the Department of Mathematics, Ewha Womans University, Seoul, South Korea.  Since October 2020, he is currently in the School of Computer Science, Nanjing University of Posts and Telecommunications, Nanjing, China, where he is currently a Professor appointed by the president.  His research interests include  coding theory and cryptographic
functions.
\end{IEEEbiographynophoto}

\begin{IEEEbiographynophoto}{Jong Yoon Hyun}  received his B.S. degree from Dongguk University in 1997
and the M.S. (2002), Ph.D. (2006) degrees in Mathematics from POSTECH.
He has been a professor currently in Konkuk University, Glocal Campus, Chungju, South Korea. His research interests include coding theory, information theory, cryptographic
functions and algebraic graph theory.
\end{IEEEbiographynophoto}

\begin{IEEEbiographynophoto}{Yoonjin Lee}  received her B.S. degree in Mathematics Education from Ewha Womans University, Korea, in 1992, her M.S. degree in Mathematics from Brown University in 1996, and her Ph.D. degree in Mathematics
from Brown University, Providence, US, in 1999 under the supervision of Professor M.I. Rosen. Since getting her doctoral degree, she has worked as a faculty member of the department of Mathematics at several universities in the US and Canada: Arizona State University (1999-2000), University of Delaware (2000-2002), Smith College (2002-2005) and Simon Fraser
University (2005-2007). She has been a professor in the department of Mathematics of Ewha Womans University since 2007, and she is a chief-in-editor of the Bulletin of the Korean Mathematical Society. Ewha Womans University is the largest women institution nationwide and it is her alma mater. Her research centers on algebraic number theory and algebraic coding theory with emphasis on the following aspects: arithmetic of function fields, Drinfeld modules, self-dual codes and cryptographic functions.
\end{IEEEbiographynophoto}





\end{document}